\documentclass[aps,prl,10pt,nofootinbib,amssymb,
amsmath,twocolumn,superscriptaddress,showpacs,showkeys]{revtex4}
\usepackage{amssymb}
\usepackage{dsfont}
\usepackage{amsmath}
\newcommand{\be}{\begin{equation}}
\newcommand{\ee}{\end{equation}}
\newcommand{\bea}{\begin{eqnarray}}
\newcommand{\eea}{\end{eqnarray}}

\newcommand{\pro}{\partial}

\newcommand{\hA}{{\hat A}}

\newcommand{\ba}{\begin{array}}
\newcommand{\ea}{\end{array}}

\newcommand{\nn}{\nonumber}
\newcommand{\mn}{{\mu\nu}}

\newcommand{\Int}{\displaystyle\int}

\begin{document}

\title  {On gauge invariant nucleon spin decomposition}
\bigskip

\author{P. M. Zhang}
\affiliation{ Institute of Modern Physics, Chinese Academy of Sciences,
 Lanzhou 730000, China }
  \affiliation{ Research Center for Hadron and CSR Physics,
 Lanzhou University \& Institute of Modern Physics of CAS, Lanzhou 730000, China}
\author{D. G. Pak}
\affiliation{ Institute of Modern Physics, Chinese Academy of Sciences,
 Lanzhou 730000, China }
\affiliation{Bogoliubov Laboratory of Theoretical Physics, Joint Institute for Nuclear Research, \\
         Dubna, Moscow region, 141980, Russia }

\begin{abstract}
A non-uniqueness problem of gauge invariant separation of quark and gluon
contributions to nucleon spin is considered. We show that there is a wide number
of gauge invariant spin decompositions each of them reduces to the
canonical one in a special gauge. A class of physical gauge equivalent nucleon
spin decompositions is selected by requirements of consistence with
helicity notion described within $E(2)$ little group representation theory
and with gluon helicity $\Delta g$ measured in experiment.
\end{abstract}
\vspace{0.3cm}
\maketitle

It has been a long standing problem of gauge invariant
definition of gluon spin and orbital angular momentum
\cite{JM, ji}. Recently a gauge invariant decomposition of
the total nucleon angular momentum into quark and gluon constituents
has been proposed \cite{chen1,chen2}, and subsequently other possible
gauge invariant decompositions for nucleon spin
have been suggested \cite{wakam1,wakam2,cho1,cho2,hatta}.
Despite this progress there are still
principal controversies on fundamental conceptual level
in determining a consistent notion for gluon spin
and orbital angular momentum \cite{ji2,leader}.
In this Letter we revise the problem of nucleon spin
decomposition and existence of a consistent gauge invariant
concept of spin in the non-Abelian gauge theory.

Let us start with the well known canonical decomposition of nucleon
total angular momentum in quantum chromodynamics (QCD)
\bea
&&  J^{can}_{\mu \nu}= \Int {\bar \psi} \gamma^0 \dfrac{\Sigma_\mn}{2} \psi d^3 x
-i \Int {\bar \psi} \gamma^0  x_{[\mu} \pro_{\nu]} \psi d^3 x \nn\\
&&- \Int \vec A_{[\mu} \cdot \vec F_{\nu] 0} d^3 x
- \Int \vec F_{0 \alpha} \cdot x_{[\mu} \pro_{\nu]} \vec A_{\alpha} d^3 x, \label{canon}
\eea
where we use vector notations for vectors in color space.
All terms in this decomposition, except the first one, are not gauge invariant.
In the series of papers \cite{chen1,chen2,chen3} Chen et al have proposed
a gauge invariant decomposition of the total angular momentum in
quantum electrodynamics (QED) and QCD.
The basic idea in Chen et al approach is to separate pure gauge and physical
degrees of freedom of the gauge potential in a gauge covariant way.
Let us rewrite the canonical angular momentum
using a more general split of the gauge potential into two independent
parts
\bea
\vec A_\mu=\vec B_\mu+\vec Q_\mu, \label{split}
\eea
where the field $\vec B_\mu$ is a so-called background (classical) field,
and $\vec Q_\mu$ is an analog of the quantum field in the framework
of the covariant background quantization formalism \cite{dewit,honerkamp}.
Notice, the background field $\vec B_\mu$ transforms as a gauge potential
whereas the quantum field $\vec Q_\mu$ transforms as a covariant color
vector under the classical type of gauge transformation \cite{dewit,honerkamp}.
Such a general form of splitting is useful in constructing nucleon
spin decomposition schemes with dynamic quark momentum \cite{cho1,cho2,dspin}.
Adding a surface term
\bea
\int d^3 x \pro_\alpha (\vec F_{0\alpha} \cdot x_{[\mu}\vec B_{\nu]})
\eea
to the canonical angular momentum (\ref{canon}), and using the equation of motion
\bea
&& D^\mu \vec F_{0\mu}=i \bar \psi \gamma^0 \psi
\eea
one can obtain the following expression for the total angular momentum tensor
\bea
&& J_{\mu\nu}^{can}=\nn \\
&&\int d^3 x \Big \{ \bar \psi \gamma^0 \dfrac{\Sigma_{\mu\nu}}{2} \psi-
i \bar \psi \gamma^0 x_{[\mu} {\cal D}_{\nu]}\psi-\vec F_{0[\mu}\cdot \vec Q_{\nu ]}\nn \\
&&-\vec F_{0 \alpha}\cdot x_{[\mu}({\cal D}_{\nu]}
             \vec Q_\alpha-\vec{\cal F}_{\nu ]\alpha}
             \mbox{\small $(B)$}) \Big \}, \label{canoncov}
\eea
where ${\cal D}_\mu$ contains the background field $\vec B_\mu$,
and $\vec{\cal F}_{\nu\alpha}\mbox{\small $(B)$}$ is a field strength defined in terms
of $\vec B_\mu$ only.
The given expression for the total angular momentum is quite general, and it is
valid irrespectively of further imposing any constraints
on the fields $\vec B_\mu, \vec Q_\mu$. In particular,
one can identify the field $\vec B_\mu$ with
a pure gauge field $\vec A_\mu^{pure}$ and the field $\vec Q_\mu$ with a
physical gauge potential $\vec A_\mu^{phys}$
by imposing two conditions
\bea
&&\vec {\cal F}_{\mu\nu} \mbox{\small $(A^{pure})$}=0,~~~~~~~\nn \\
&& {\cal D}_i\vec A_i^{phys}=0, \label{condition2}
\eea
where Latin letters are used for space-like indices.
With this the spin decomposition (\ref{canoncov})
reproduces the gauge invariant decomposition for the nucleon angular momentum
proposed in \cite{chen2}.
It has been shown that the constraints (\ref{condition2}) can be solved
by perturbation theory producing a solution for $\vec A_\mu^{phys},\vec A_\mu^{pure}$
in terms of the unconstrained gauge potential \cite{chen3}.
Notice, explicit solutions for
$\vec A_\mu^{phys},\vec A_\mu^{pure}$ represent
non-local functionals of the initial gauge potential $\vec A_\mu$.
The final expression for the gauge invariant decomposition of nucleon spin
corresponds to the space vector part of the Eqn. (\ref{canoncov}) \cite{chen2}.
Each term in the decomposition has become gauge invariant due to covariant
transformation law
for the physical field $\vec A^{phys}$.
In the gauge $\vec A^{pure}=0$ the decomposition
reduces to the canonical one in the standard Coulomb gauge.
Even though the given decomposition is gauge invariant,
it is not satisfactory since the basic constraint (\ref{condition2})
defining the notion of the physical gauge potential $\vec A_\mu^{phys}$
is not Lorentz invariant.
This might imply that the matrix elements of the spin density operator and
orbital angular momentum will be frame dependent. Another serious problem is
whether such a decomposition of nucleon spin is unique.

The gauge and Lorentz invariant nucleon spin decomposition
has been suggested recently in \cite{cho1}. The defining equation
for the physical field is given by the constraint of Lorenz gauge type
\bea
&& {\cal D}^\mu\vec A_\mu^{phys}=0. \label{Lorcon}
\eea
There are other known Lorentz invariant gauge conditions, the Gervais-Neveu and
Fock-Schwinger gauges.
Notice, that the Fock-Schwinger gauge $x^\mu A_\mu=0$ allows to express the
vector potential in terms of the field strength in a simple way\footnote[1]{
private communication from
W.M. Sun, F. Wang suggesting the spin decomposition with Fock-Schwinger type constraint}
 \cite{Smilga:1982wx}
\bea
&& A_\mu=\int_0^1 d \alpha \alpha x^\nu F_{\nu\mu}(\alpha x).
\eea
Unfortunately, it lacks the invariance under translations.
So that, the Lorenz gauge type constraint is a unique gauge which satisfies
the Poincare and conformal symmetries. A solution to the constraint (\ref{Lorcon})
for $\vec A_\mu^{phys}$ in terms of unconstrained gauge potential can be obtained by
perturbation method in a similar manner as in the case of Chen et al decomposition \cite{chen3}.
However, solving the Lorenz gauge constraint (\ref{Lorcon}) on mass-shell, i.e.,
on the space of solutions to equations of motion, encounters a serious problem.
The problem becomes evident in the case
of Maxwell theory where the formal solution is given by
\bea
 A_\mu^{phys}(\vec x,t)=\int d^3 \vec x'
\dfrac{\pro^\nu F_{\nu\mu}(x',t-|\vec x-\vec x'|/c)}{|\vec x-\vec x'|}.
\eea
In the case of absence of matter fields the r.h.s. of the equation
vanishes identically due to equations of motion, so that the
physical field $A_\mu^{phys}$ can not be determined. This is a well-known consequence
of the incompleteness of the Lorenz gauge. So that one has to
impose an additional condition to provide the transversality property of the
real photon (gluon). One may choose, for instance,
the Coulomb gauge constraint and solve it as Chen et al did,
but then we will return to the problem of Lorentz frame dependence.
The problem becomes worse since the choice of Lorentz non-invariant constraint
for $\vec A_\mu^{phys}$ is not unique unlike the case of Lorenz gauge condition.

Each spin decomposition corresponding to a special constraint for
physical field defines a class of gauge equivalent
operators $\{\vec A_\mu^{phys}\}$.
Since the physical field $\vec A_\mu^{phys}$ is gauge covariant
it can be expressed in terms of the field strength and its
covariant derivatives. However, it is known  that for a given
field strength there may exist gauge nonequivalent potentials
\cite{math1,math2}. This implies that various $\vec A_\mu^{phys}$
given as solutions to different physical constraints
will lead to gauge nonequivalent operators and,
in general, to different matrix elements.
To select which class of gauge equivalent physical fields
$\vec A_\mu^{phys}$  produces a
proper spin operator we will require the
consistence of definitions for $\vec A_\mu^{phys}$
with helicity notion described in the framework of group
representation theory.
We will demonstrate this by explicit constructing two different
spin decompositions based on using gauge invariant variables
and generalized axial gauge type constraint for the physical gauge field.
Finally we will prove the relation
between the gauge invariant definition of gluon spin density
and gluon helicity $\Delta g$.

Let us first consider a decomposition scheme based on the known notion
of gauge invariant variables in $SU(2)$ Yang-Mills theory
\cite{perv1}. Main definitions of the gauge invariant variables can be
generalized straightforward to the case of $SU(3)$ QCD.
The main idea in constructing gauge invariant variables
is to find a pure gauge $SU(3)$ matrix field in terms
of the initial gauge potential $\vec A_\mu$.
The key observation is that the gauge transformation of the
temporal component $A_0^a, ~ (a=1,2,...,8)$ is given
by a covariant time derivative of the gauge parameter. So that,
$A_0^a$ can be expressed in a pure gauge form
in terms of $SU(3)$ matrix valued function ${\mathbf{v}}$
\bea
&& \hat A_0={\mathbf{v}}^{-1} \pro_0 {\mathbf{v}}, \label{eqA0}
\eea
where $\hA_\mu \equiv gA_\mu^a \dfrac{\lambda^a}{2i}$,
and $\lambda^a$ are Gell-Mann matrices.
Notice, the matrix function ${\mathbf{v}}$ transforms
covariantly under gauge transformation, ${\mathbf{g}}\in SU(3)$,
\bea
{\mathbf{v}}(A^{\mathbf{g}})={\mathbf{v}}{\mathbf{g}}^{-1}.
\eea
Using Eqn. (\ref{eqA0}) and the equation of motion for the temporal component $A_0^a$
\bea
 (D_i^2 A_0)^a&=&(D_i \pro_0 A_i)^a - j_0^a, \nn  \\
j_0^a&=&g \bar \psi \gamma^0 \dfrac{\lambda^a}{2} \psi,\label{eqmot1}
\eea
one can write down the equation defining the matrix
function ${\mathbf{v}}(A)$
\bea
&& \pro_0 {\mathbf{v}}(A) = {\mathbf{v}}(A) \Big ( \dfrac{1}{D_i^2(A)}
 [D_j (A) \pro_0 \hA_j-\hat j_0] \Big ), \label{eqP1}
\eea
where the equation may include or not the source term.

The solution to the equation (\ref{eqP1})
can be obtained in the form of time ordered exponent \cite{perv1}
\bea
{\mathbf{v}}(A)=T \exp \Big {\{} \int^t dt \Big (\dfrac{1}{D_i^2(A)} [D_j(A) \pro_0 \hA_j-\hat j_0]
 \Big ) \Big {\}}.    \label{Texp}
\eea
This allows to define gauge invariant variables $\hA_i^I(A)$ and $\psi^I(A,\psi)$
in terms of the original gauge potential and matter fields \cite{perv1}
\bea
\hA_i^I(A)&=& {\mathbf{v}}(A) (\pro_i+\hA_i) {\mathbf{v}}^{-1}(A), \nn \\
\psi^I(A,\psi)&=&{\mathbf{v}}(A) \psi.    \label{GIvars}
\eea
One can check that $\hA_i^I$ satisfies a constraint which represents a
generalized covariant Coulomb gauge condition
\bea
D_i(A^I) \pro_0 \hA_i^I-\hat j_0 =0. \label{covCG}
\eea
One should stress, that we do not impose this condition,
it is fulfilled identically due to definitions of $\hA_i^I$ and ${\mathbf{v}}(A)$.
Since $\hA_i^I$ and Eqn. (\ref{covCG}) are gauge invariant,
the simplest way to prove the identity (\ref{covCG})
is to consider the relationship (\ref{Texp}) in a special gauge ${\mathbf{v}}(A)=\mathbf{1}$.
Finally, the first equation in (\ref{GIvars}) can be
written in the following inverted form
\bea
\hA_i&=&{\mathbf{v}}^{-1}(A) \pro_i {\mathbf{v}}(A)+ {\mathbf{v}}^{-1}(A) \hA_i^I(A) {\mathbf{v}}(A)
               \nn \\
 &\equiv& \hA_i^{pure}+\hA_i^{phys},  \label{splitP}
\eea
where we can identify the first and the second term as pure gauge and physical
gauge potentials needed to construct a desired gauge invariant spin decomposition.
The temporal components of the pure gauge and physical fields are defined as follows
\bea
&&\hat A_0^{pure}={\mathbf{v}}^{-1}(A) \pro_0{\mathbf{v}}(A),\nn \\
&&\hat A_0^{phys}=\hat A_0-\hat A_0^{pure}. \label{Aphys}
\eea
The corresponding total angular momentum decomposition
is given by the Eqn. (\ref{canoncov}) with the replacement
$\vec B_\mu \leftrightarrow  \vec A_\mu^{pure},~
\vec Q_\mu\leftrightarrow  \vec A_\mu^{phys}$.
The decomposition reduces to the canonical one in the gauge
${\mathbf{v}}(A)=\mathbf{1}$, which implies $\vec A_\mu^{phys}=\vec A_\mu$ and the constraint
(\ref{covCG}) turns into a generalized Coulomb gauge condition for the gauge potential $\vec A_i$.
Quantization with the constraint (\ref{covCG}) in Hamiltonian formalism
had been done in \cite{fried}. The consistency of the gauges depending on
time had been proved in \cite{barb}.
One should notice that the pure gauge field ${\mathbf{v}}(A)$ in the absence of dynamic gluons describes
also pure gauge Gribov modes which can be separated explicitly \cite{perv1}.
Due to that, there is no Gribov ambiguity problem in the presented decomposition.

The important feature of the spin decomposition based on the gauge invariant variables
is that due to equations (\ref{eqmot1}, \ref{eqP1})
the time component of the physical field defined in (\ref{Aphys})
vanishes, $\vec A_0^{phys}=0$, before imposing any gauge fixing condition.
This allows to prove that the decomposition
satisfies the requirement of consistence with the helicity notion
from the point of view of group representation theory.
Notice, that due to the condition $\vec A_0^{phys}=0$
the constraint (\ref{covCG}) in free space can be written
explicitly as the transversality condition for
the color electric field $\vec E_i^I$
\bea
&&D_i(A^I)\vec E_i^{I} =0,~~~~~ \vec E_i^{I}=\pro_0 \vec A_i^{I}.
\eea
The transversality for the physical gauge potential $\vec A^{phys}$
will be implied naturally from the helicity gauge conditions considered
below. The only frame independent notion of spin in the gauge theory
for a massless particle is the helicity which can be described
within the framework of the little group $E(2)$ of the Lorentz group \cite{wigner}.
The gauge invariant consideration of the helicity in QED had been done
in \cite{weinberg}. We will generalize consideration of helicity
notion to the case of non-Abelian gauge theory following
the non-covariant treatment of the problem in a similar way as it had been done
in Abelian Maxwell theory \cite{kim}. The construction of physical gauge potential
$\vec A_\mu^{phys}$ plays a crucial role.
If the gluon momentum is directed along the $z$ axis, $p_\mu=(\omega,0,0,\omega)$,
the generators of the little group $E(2)$ are given by the rotation generator $J_3$,
which is the helicity operator in this case,
and by combinations of boost and rotation generators:
\bea
&&J_3,~~~N_1=K_1-J_2,~~~~~N_2=K_2+J_1.
\eea
By definition the transformations of the
little group $E(2)$ leave the gluon momentum invariant.
For the gauge potential $\vec A_\mu^{phys}$ to represent helicity eigenstates
of the operator $J_3$ one must have the so-called
helicity gauge conditions \cite{kim}
\bea
&& \vec A_0^{phys}=0, ~~~~~~~
\vec A_3^{phys}=0.  \label{helicity}
\eea
To provide both helicity conditions in a consistent manner with equations of motion
has been a principal obstacle toward generalization to the case of non-Abelian gauge theory.
In our approach, since one has the condition $\vec A_0^{phys}=0$ on mass-shell by construction,
it is possible to provide the second helicity condition $\vec A_3^{phys}=0$
by choosing a gauge of either Coulomb or axial or light-cone type.
This is the main result which allows to select a physical gauge covariant
operator $\vec A^{phys}(A)$ and corresponding spin density
consistently with the helicity notion.

Now, it becomes clear that there should exist a class of gauge equivalent
spin decompositions which satisfy on mass-shell the same helicity conditions
for the physical field.
One such possible decomposition has been proposed recently \cite{hatta}.
Let us consider other decomposition schemes with a generalized axial gauge
type constraint for the physical field and without invoking the concept of an invariant field.
One can construct such decompositions
in a surprisingly simple way. Let us define the physical gauge potential $\vec A_\mu^{phys}$
by a generalized axial gauge type constraint
\bea
n^\mu \vec A_\mu^{phys}=0, \label{l-c}
\eea
where the vector $n^\mu$ specifies the axial or light-cone gauge condition.
Notice, that the defining equation (\ref{l-c}) for
$\vec A_\mu^{phys}$ is similar to a generalized axial gauge condition
for the gauge potential used in QCD which admits an explicit solution
in terms of the field strength
\cite{bal}. This allows to write down the expression for the
physical gauge field $\vec A_\mu^{phys}$ in terms of the general field strength
as follows
\bea
&& \vec A_\mu^{phys}=-\int_0^\infty d\lambda n^\nu \vec F_{\nu\mu}(x+\lambda n). \label{AF}
\eea
A pure gauge field $\vec A_\mu^{pure}$ is defined by
\bea
&& \vec A_\mu^{pure}=\vec A_\mu- \vec A_\mu^{phys}.\label{AF2}
\eea
One can check that $\vec A_\mu^{pure}$ satisfies the pure gauge condition $\vec F_{\mu\nu}^{pure}=0$.
By choosing the proper vector $n_\mu$ one can define the physical gauge
potential $\vec A_\mu^{phys}$ by choosing either axial, $\vec A_3^{phys}=0$, or light-cone,
$\vec A_+^{phys}=0$, equation.
The both helicity gauge conditions (\ref{helicity})
can be easily reached by imposing the temporal
gauge fixing condition $\vec A_0^{phys}=0$.
The advantage of the decomposition with
the light-cone type constraint (\ref{l-c}), $n^2=0$,
is that the corresponding non-local operator $\vec A_\mu^{phys}(A)$
reduces to the canonical spin density operator in a special
gauge $\vec A_\mu^{pure}=0$, i.e., explicitly in the light-cone gauge. This
allows to make straightforward one-to-one correspondence of the
gauge invariant spin density operator
to the gluon helicity $\Delta g$ measured in experiment.
Notice, that gauge invariant spin density operator
in other known spin decomposition schemes reduces to the canonical one
in a different gauge.
Let us write the four-vector for gluon spin operator
corresponding to the canonical gluon spin density
\bea
S_\mu^{gluon}=\epsilon_{\mu\nu\rho\sigma}\vec F_{\nu\rho} \cdot \vec A_\sigma^{phys}. \label{spinvec}
\eea
Substituting (\ref{AF}) into the last equation
one can express the spin vector in light-cone gauge
in a similar manner as in \cite{bal}
\bea
&&S_\mu^{gluon}=Tr \int_0^\infty d\lambda n^\nu F_{\nu \xi}(\lambda n+z)
P\exp \Big (i g \cdot \nn \\
&& \int_0^\lambda du n^\nu A_\nu (un+z) \Big ) \tilde F_{\xi\mu}(z)
 +n_\mu ({\cal O}(A^3)), \label{PK}
\eea
where $\tilde F_{\xi\mu}^a$ is the dual field strength, and
we have omitted the terms coming from the cubic terms in (\ref{spinvec})
proportional to $n_\mu$,
since they will not contribute to the matrix element of $n_\mu S_\mu$
due to light-cone condition $n^2=0$.
On the other hand, one has a simplified
expression for the gluon helicity  at light-cone
$x^2=0$ \cite{manohar,jaffe}
\bea
(sx) \Delta g=<N|\int_0^\infty d\lambda x^\mu F_{\xi\mu}(\lambda x) \cdot \nn \\
P\exp \Big (i g \int_0^\lambda du x^\nu  A_\nu (ux) \Big ) x^\nu \tilde F_{\nu\xi}(0)|N>,
\eea
where $s_\mu=\bar u(p,s)\gamma_\mu\gamma_5 u(p,s)$ is the four-vector of nucleon
spin. With (\ref{PK})
one results in the known relationship between $\Delta g$ and
the nucleon expectation value of the transverse part of $S_\mu$
\bea
&&<N|x^\mu S_\mu^{gluon}|N>=-(sx)\Delta g.
\eea

In conclusion, there is a wide number of possible
gauge invariant spin decompositions suggested in \cite{chen1,chen2,wakam1,wakam2,cho1,cho2,hatta}
and in the present paper. In general they lead to
gauge nonequivalent gluon spin operators. The Poincare and conformal
invariance selects a unique Lorentz invariant decomposition
with the Lorentz type constraint for physical field \cite{cho1,cho2}.
However, since this decomposition
is not well defined on mass-shell, its physical meaning is unclear.
For most of Lorentz non-invariant decompositions the definition
of spin operator is frame dependent. We
have shown that there is a class of gauge equivalent spin decompositions,
(\ref{splitP},~\ref{AF}), leading to gauge invariant gluon spin operators
consistent with the helicity notion and, so that, such definitions
of spin operators are frame independent. In general, the gauge invariant spin densities
represent spatially and temporally non-local functionals. However,
this non-locality has no physical meaning since it can be removed in a
special gauge resulting in a standard local expression for
the canonical spin decomposition.
In practical use the decomposition defined by (\ref{AF},~\ref{AF2}) is more suitable.
The corresponding definitions for the spin operator are gauge equivalent
and lead to the same matrix elements. Notice, that the notion of
such defined gauge invariant spin operator with using a non-local
operator function for the physical field $\vec A_\mu^{phys}$
is similar to the notion of
the quantum effective action which is gauge invariant but
has a different operator form depending on a chosen gauge.

As it is known, there are two principal issues in the nucleon
spin decomposition problem. The first one is how to separate
contributions of nucleon constituents.
Another one is related to the problem of observability in experiment.
In QED the photon spin and orbital angular momentum
are measurable quantities \cite{exp1,exp2,exp3,exp4,exp5},
and they correspond to the canonical decomposition \cite{theor1,theor2,theor3,theor4,theor5}.
In QCD, since hadrons represent strongly bound states,
other schemes of nucleon spin decomposition with
dynamic quark momentum  might be more relevant \cite{wakam1,wakam2,cho1,cho2,dspin}.
This problem will be considered in a separate paper \cite{CAS1L}.

{\bf  Acknowledgements}

One of authors (DGP) thanks N.I. Kochelev, A.E. Dorokhov, I.V. Anikin, O.V. Teryaev, A.V. Efremov
for numerous discussions and hospitality during his staying in BLTP, JINR.
We thank Y.M. Cho for critical comments and
F. Wang, W.M. Sun for numerous interesting discussions.
The work was supported by NSFC Grants (Nos. 11035006 and 11175215),
and by CAS (Contract No. 2011T1J31).

\end{document}